\documentclass[pra,twocolumn]{revtex4}

\usepackage{graphicx,epsfig}
\usepackage{amssymb,amsmath}

    \newcommand{\g}{\gamma}    
\newcommand{\e}{\epsilon}     

\renewcommand{\a}{\alpha}

\renewcommand{\th}{\theta}   
\renewcommand{\d}{\delta}    
\renewcommand{\l}{\lambda}

\renewcommand{\S}{\Sigma}       

\renewcommand{\o}{\omega}       

\newcommand{\hf}{\tfrac{1}{2}}    

\newcommand{\ord}{\mathcal{O}}
\newcommand{\ra}{\rightarrow}

\newcommand{\lba}{\left(}    \newcommand{\rba}{\right)}
\newcommand{\lbc}{\left[}    \newcommand{\rbc}{\right]}
\newcommand{\lbb}{\left\{}    \newcommand{\rbb}{\right\}}

\newcommand{\bra}[1]{\langle\left.{#1}\right|}
\newcommand{\ket}[1]{\left|{#1}\right.\rangle}
\newcommand{\xpct}[1]{\langle{#1}\rangle}    % expectatn value

\newcommand{\vp}{{\bf p}}  % usual vector quantities
\newcommand{\vq}{{\bf q}}  % double bracketing not required with \vec
\newcommand{\vk}{{\bf k}}  % but required with \bf
\renewcommand{\vr}{{\bf r}}

\newcommand{\hc}{\hat{c}}  \newcommand{\hcd}{\hat{c}^\dag} 

\newcommand{\hD}{\hat{D}}  \newcommand{\hDd}{\hat{D}^\dag} 
\newcommand{\hS}{\hat{S}}  \newcommand{\hSd}{\hat{S}^\dag} 
\newcommand{\hX}{\hat{X}}  \newcommand{\hY}{\hat{Y}}

\newcommand{\hb}{\hat{b}}
\newcommand{\hbd}{\hat{b}^\dag}

\newcommand{\heta}{\hat{\eta}}
\newcommand{\hetad}{\hat{\eta}^\dag}
\newcommand{\hph}{\hat{\phi}}
\newcommand{\hphd}{\hat{\phi}^\dag}
\newcommand{\nG}{n_{\rm G}}

\renewcommand{\ss}{s}
\newcommand{\ch}[1]{\cosh(\ss_{#1})}
\newcommand{\sh}[1]{\sinh(\ss_{#1})}
\newcommand{\Nc}{N_{\rm c}}
\newcommand{\nc}{n_{\rm c}}
\newcommand{\wt}{\widetilde}

\begin{document}

\title{Squeezing in the weakly interacting Bose condensate}

\author{Masudul Haque}

\affiliation{Institute for Theoretical Physics, Utrecht University,
The Netherlands}

\author{Andrei E.~Ruckenstein}

\affiliation{BioMaPS Institute for Quantitative Biology and Department
  of Physics and Astronomy, Rutgers University, Piscataway, NJ, USA}

\date{\today}

\begin{abstract}

We investigate the presence of squeezing in the weakly repulsive
uniform Bose gas, in both the condensate mode and in the nonzero
opposite-momenta mode pairs, using two different variational
formulations.  We explore the $U(1)$ symmetry breaking and Goldstone's
theorem in the context of a squeezed coherent variational
wavefunction, and present the associated Ward identity.  We show that
squeezing of the condensate mode is absent at the mean field
Hartree-Fock-Bogoliubov level and emerges as a result of fluctuations
about mean field as a finite volume effect, which vanishes in the
thermodynamic limit. On the other hand, the squeezing of the
excitations about the condensate survives the thermodynamic limit and
is interpreted in terms of density-phase variables using a
number-conserving formulation of the interacting Bose gas.

%\pacs{03.75.Hh 05.30.Jp}
\end{abstract}

\maketitle

\section{Introduction}   \label{sect_intro}

The quantum optics concepts of minimum-uncertainty states such as
coherent and squeezed states have been applied to quantum
condensed-matter systems in a variety of settings.  The study of the
Bose gas with weak repulsive interactions has benefited greatly from
borrowing and extending the ideas of coherent states originally
developed by Glauber \cite{glauber63a, glauber63b} in optics,
particularly in the context of the Bose-Einstein condensate
\cite{gross60, CummingsJonhston66, langer68,
langer69, BarnettBurnettVaccarro96, Valatin_BoulderLecture63}. In the traditional treatments of
the interacting Bose gas, pairs of opposite momentum excitations can be
created (destroyed) out of (into) the condensate, and one can
interpret this effect in terms of squeezing of pairs of
opposite-momentum excitations induced by inter-particle interactions.
Furthermore, squeezing within the condensate mode itself is another
intriguing question, related to higher moments of the condensate-mode
annihilation and creation operators \cite{navez98, SolomonFengPenna99,
  DunninghamCollettWalls98, RS-Choi-New-Burnet02:quantumstate,
ValatinButler_NC58, Valatin_BoulderLecture63, GlassgoldSauermann69a,
GlassgoldSauermann69b}. Thus, naively the interacting Bose gas
displays two types of squeezing super-imposed on properties of a
coherent state that would, strictly speaking, only represent the
behavior of a non-interacting gas. While "squeezing" effects have been
known in some form or other for a long time, it is only recently that
the language of minimum-uncertainty states has been used in
descriptions of the Bose gas \cite{navez98, SolomonFengPenna99,
DunninghamCollettWalls98, ChernyakChoiMukamel03,
RS-Choi-New-Burnet02:quantumstate}.
Despite several studies, there remain significant questions concerning
the existence and physical interpretation of squeezing in the Bose gas
ground state.
In this Article, we seek insight into the intuitive physical meaning
of squeezing in the context of the weakly interacting Bose gas,
treating both kinds of squeezing mentioned above: the single-mode
squeezing within the zero-momentum condensate mode, as well as the
pair-wise squeezing of finite-momenta bosons.

A coherent state encodes the physics of having a definite phase at the
expense of strict number conservation, which to the condensed matter
community is the essence of Bose condensation.  (See however
Refs.~\cite{fixed-N:CastinDum98, fixed-N:LewensteinYou96,
fixed-N:IlluminatiNavezWilkens99, fixed-N:gardiner97, fixed-N:AER00,
girardeau_gardiner-note} for attempts to circumvent number
conservation violation.)  As a result coherent states were appreciated
early in the study of the Bose condensate \cite{gross60,
CummingsJonhston66, langer68, langer69, Valatin_BoulderLecture63}.  In
addition, the need to incorporate $\pm\vk$ correlations also led to
squeezing operators similar to
$\exp\lbc\g\hcd_{\vk}\hcd_{-\vk}-\g^*\hc_{\vk}\hc_{-\vk}\rbc$, which
today would be called ``squeeze'' operators, being used for the
interacting Bose gas since the 1960s \cite{gross60,
ValatinButler_NC58, Valatin_BoulderLecture63, CummingsJonhston66,
GirardeauArnow59}.  In addition, some early authors have also
incorporated single-mode squeezing explicitly in the condensate mode
itself \cite{ValatinButler_NC58, Valatin_BoulderLecture63,
GlassgoldSauermann69a, GlassgoldSauermann69b}.
%
%Squeezing in the condensate mode will be studied further in this
%Article.

During the resurgence of interest in the interacting Bose gas in the
1990s, several studies on squeezing in the Bose gas have been
performed explicitly using the quantum-optics language now available.  
Studies of the quantum state of \emph{trapped} condensates
\cite{DunninghamCollettWalls98, RS-Choi-New-Burnet02:quantumstate}
have indicated the presence of squeezing in the condensate mode
itself, which corresponds to $\vk=0$ mode squeezing in the uniform
case.  
Ref.~\cite{ChernyakChoiMukamel03} uses a ``generalized coherent
state'' (including both single-mode and two-mode squeezing), to derive
time-dependent Hartree-Fock-Bogoliubov equations for a non-uniform
Bose gas.
%
%% Ref.~\cite{ChernyakChoiMukamel03} describes a non-uniform Bose gas
%% using a ``generalized coherent state'' containing coherent populations
%% in each mode as well as both single-mode and two-mode squeezing.  The
%% derivation of the time-dependent Hartree-Fock-Bogoliubov equations
%% \cite{BogoldeGennes} using this formulation may be regarded as an
%% extension of using a pure coherent state to derive the
%% Gross-Pitaevskii equation \cite{langer68, langer69,
%% BarnettBurnettVaccarro96}.
%
Refs.~\cite{navez98} and \cite{SolomonFengPenna99} have both used a
wavefunction containing a squeezed coherent state for the condensate
mode, and the usual $\pm\vk$ pair squeezed vacua for the $\vk{\ne}0$
modes, similar to our first variational wavefunction $\ket{\rm sq1}$
in Sec.~\ref{sect_formalism}.
Ref.~\cite{navez98} considers squeezing in the condensate mode (as we
do in Secs.~\ref{sect_formalism} and \ref{sect_U1}) and focuses on 
regulating anomalous fluctuations, 
%
%by using linear combinations of the states with different phases.  
%
while Ref.~\cite{SolomonFengPenna99} uses the formalism to calculate
coherence functions \cite{glauber63a, loudon00_B}.   
%to compare with cold-atom experiments.

Our main results are as follows.  We find that the condensate mode is
indeed squeezed, but the scaling of the squeeze parameter with system
size is such that squeezing has no thermodynamic effects.  For
finite-size systems, the presence of appreciable squeezing is
determined by the competition of two small parameters.
% 
%$an^{1/3}$ and $N^{-2/3}$.  
%
For Bose-Einstein condensates in traps, this is the same competition
that determines whether the density profile is gaussian or is given by
the Thomas-Fermi approximation.
We have also formulated the Hugenholz-Pines (H-P) theorem
\cite{HP59,HohMart64} in the context of our variational formulation.
The H-P theorem enforces the absence of a gap in the excitation
spectrum of the condensed Bose gas.  (For a modern description, see,
e.g., Ref.~\cite{ShiGriffin98}).  We use the H-P theorem, or the
equivalent requirement of gaplessness, to prove that any
condensate-mode squeezing present in the system must come from beyond
a mean-field treatment of the theory.

In addition, we give a physical interpretation to the pair-wise
squeezing of boson pairs induced by condensate depletion, by using an
alternate variational state, based on the number-conserving
formulation of the Bose-condensed state in Ref.~\cite{fixed-N:AER00}.
The finite-momenta squeezing can be expressed in a ``quadrature''
space of density-oscillation and phase operators.  The squeezing of
fluctuations is found to be in the density-oscillation direction.

The paper is organized as follows.  In Sec.~\ref{sect_formalism}, we
briefly review relevant concepts of coherent and squeezed states
(Sec.~\ref{sect_quantopt}), and then construct our first variational
wavefunction for the zero-temperature Bose gas (Sec.~\ref{sect_wf}).
%
%% A similar formulation was used in 1969 in
%% Refs.~\cite{GlassgoldSauermann69a, GlassgoldSauermann69b} and more
%% recently using quantum-optics language in Ref.~\cite{navez98,
%% SolomonFengPenna99}.
%
This is used to derive the scaling of condensate-mode squeezing
properties with system size (Sec.~\ref{sect_xpctns-n-squeezings}).  In
Sec.~\ref{sect_U1} we explore the manifestation of $U(1)$ symmetry
breaking within this formalism, formulate the relevant Ward identity
(Hugenholz-Pines theorem \cite{HP59, HohMart64, ShiGriffin98}), and
construct the excitation spectrum.  These results lead to additional
physical inferences about the condensate-mode squeezing, which are
presented in Sec.~\ref{sect_ss0-concl}.
In Sec.~\ref{sect_AER} we present a second variational state, using
density-oscillation and phase variables introduced in
Ref.~\cite{fixed-N:AER00}, and use this construction to provide a
physical interpretation of $\vk{\ne}0$ squeezing.

\section{Variational treatment of Bose gas using squeezed coherent wavefunction}
\label{sect_formalism}

The three-dimensional uniform Bose gas is described by the
Hamiltonian:
\begin{equation}  \label{eq_WIBG-Hamilt}
\hat{H} = \sum_{\vk}
\lba \e_{\vk} -\mu\rba
\hat{c}_{\vk}^\dag \hat{c}_{\vk} +
\frac{U}{2V}\sum_{\vp,\vq,\vk} \hat{c}_{\vp+\vk}^\dag
\hat{c}_{\vq-\vk}^\dag \hat{c}_{\vp} \hat{c}_{\vq}\,,
\end{equation}
where $\e_{\vk}=k^2/2\tilde{m}$ is the free-gas dispersion,
$\tilde{m}$ is the boson mass, and $\hat{c}$, $\hat{c}^\dag$ are
bosonic operators.  The interaction $U$ is taken to be
momentum-independent because at low enough temperatures only $s$-wave
scattering is important: $U = 4\pi{a}\hbar^2/\tilde{m}$ modulo an
ultraviolet renormalization term, where $a$ is the $s$-wave scattering
length.  A dimensionless measure of the interaction is $an^{1/3}$,
where $n = N/V$ is the density.

In this Section, after a lightning review of the relevant coherent and
squeezed state concepts (Sec.~\ref{sect_quantopt}), we will introduce
our first variational wavefunction $\ket{\rm sq1}$ and determine the
variational parameters by minimization (Sec.~\ref{sect_wf}), and
discuss variances and squeezing properties
(Sec.~\ref{sect_xpctns-n-squeezings}).

\subsection{Quantum States of Bosonic Systems} 
\label{sect_quantopt}

Details on minimum-uncertainty quantum states can be found in quantum
optics texts and reviews, e.g., in Refs.~\cite{loudon00_B,
LoudonKnightReview87}; we give here only a brief introduction to
squeezed and coherent states and point out some applications of
squeezing concepts in condensed-matter systems.
For a bosonic mode described by operators $\hc$, $\hcd$, one defines
hermitian ``quadrature'' operators $\hX = \tfrac{1}{2}(\hc+\hcd)$ and
$\hY = \tfrac{1}{2i}(\hc-\hcd)$, conjugate to each other,
$[\hX,\hY] = i/2$, so that the uncertainty relation
$\xpct{\d^2\hX}^{1/2}\xpct{\d^2\hY}^{1/2} \ge 1/4$ is satisfied.
Coherent and squeezed states both have minimum uncertainty.

Coherent states have equal uncertainties in the quadrature directions.
A coherent state can be constructed by applying the \emph{displacement
operator} $\hD(\a) = \exp\lbc \a\hcd- \a^*\hc \rbc$ on vacuum.  The
vacuum itself is a special case with $\a=0$.
Coherent states have circular variance profiles, centered at
$(\xpct{\hX},\xpct{\hY})$ = $(\mathcal{R}e[\a],\mathcal{I}m[\a])$.

%Squeezed states were discussed first in the 1970's as ``two-photon
%coherent states'' \cite{stoler70,stoler71, yuen76}, and formalized in
%the 1980's \cite{caves80, LoudonKnightReview87}.
%
Single-mode squeezed states are produced by the squeeze operator
$\hS = \exp \lbc \g\hcd\hcd - \g^*\hc\hc \rbc$, with  $\g = \ss\
e^{i\phi}$,
whose effect is to squeeze variance profiles in the direction
indicated by $\phi/2$ on the quadrature plane.  When applied to
coherent states, $\hS$ creates squeezed coherent states:
\[
\ket{\rm sq\_coh} = \hS \ket{\rm coh} = \hS \hD \ket{\rm vac} \, .
\]
The inverted order of operators, $\hD \hS \ket{\rm vac}$, is common in
the quantum optics literature.  The fluctuations of $\hX$, $\hY$ are the
same in this alternate form, but the expectation values differ by the
factor $F = \cosh(2\ss) + \sinh(2\ss)$.  The uncertainty contour in
the $\hX$-$\hY$ plane is elliptical rather than circular, centered at
a displaced position ($\a$ or ${\a}F$).  The uncertainties are
${\hf}e^{\pm\ss}$ along the major and minor axis directions.

\begin{figure} 
\begin{center} 
\includegraphics[width=0.3\textwidth]{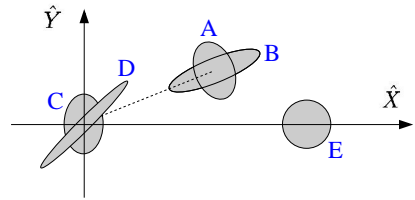}  
\caption{   \label{fig_squ_coh-en-vac} 
Fluctuation contours of squeezed coherent states (A and B), squeezed
vacua (C and D), and a coherent state with real $\a$ (E).  States A
and B have their squeeze phases locked to twice the coherence phase
($\phi=2\th$), and are therefore respectively amplitude-squeezed
($\ss<0$) and phase-squeezed ($\ss>0$) states.  State C has $\phi=0$,
$\ss<0$.  State D has larger $\ss$, and $\phi=\pi/2$, so that the tilt
is $\pi/4$.
}
\end{center}
\end{figure}

Applying the squeeze operator $\hS$ on a vacuum state produces a
single-mode squeezed vacuum, $\ket{\rm sq\_vac(single)} = \hS\ket{\rm
vac}$, which is another minimum-uncertainty state with distorted
variances in the $\hX$-$\hY$ plane, but the quadrature expectations
are now zero (states C, D in Fig.~\ref{fig_squ_coh-en-vac}).  We will
be more interested in \emph{mixed-mode} squeezed vacua,
\[
\ket{\rm sq\_vac(mixed)} = \exp \lbc \g\hcd_m\hcd_n - \g^*\hc_n\hc_m
\rbc \ket{\rm vac} \, ,
\]
which will be the fate of $\vk\ne{0}$ states in our variational
treatment of the non-ideal Bose gas.
Squeezing of uncertainty is now seen not in the space of the
individual mode quadratures, ($\hX_m$, $\hY_m$) or ($\hX_n$, $\hY_n$),
but in the \emph{mixed} quadrature variables $\hX =
\frac{1}{\sqrt{2}}(\hX_m+\hX_n$), $\hY =
\frac{1}{\sqrt{2}}(\hY_m+\hY_n$).

After squeezed states became popular in quantum optics in the
1980s, the concept was utilized in the analysis of several
condensed matter systems.
Squeezed coherent states have been used to treat variationally the
``spin-boson'' model that arises in connection with dissipative
tunneling \cite{disspn:Leggett-etal_RMP87}, defect tunneling in
solids, the polaron problem, etc.~\cite{squ_SpinBoson:ChenZhangWu89-2,
squ_SpinBoson:StolzeMuller90, LoManousakisSollieWang94,
squ_SpinBoson:DuttaJayannavar94}.  Squeezed states have also been used
for polaritons \cite{Squ-polariton:ArtoniBirman90}, exciton-phonon
systems \cite{Squ-excphon:SonnekEiermannWagner95}, many-body gluon
states \cite{SquGluon:Blaschke-etal_97}, bilayer quantum Hall systems
\cite{Squ-bilayerQHE:NakajimaAoki97}, phonon systems
\cite{SquPhonon:HuNori96}, and attractive Bose systems on a lattice
\cite{SaDeMelo91}.

\subsection{Variational wave function and minimization}  \label{sect_wf}

For a uniform condensate, the macroscopic occupation is in the
zero-momentum state, so we will use a coherent occupation of the
$\vk=0$ mode only: $\hD = \hD_0 = \exp\lbc \a\hcd_0 - \a^*\hc_0 \rbc$,
with coherence parameter $\a = f{\ }e^{i\th}$.  Intuitively, $\a$
corresponds to the order parameter for Bose condensation.

We will apply a mixed-mode squeeze operator for each opposite-momenta
mode pair.  Thus the variational ground state is $\ket{\rm sq1.gr} =
\hS\ket{\rm coh} = \hS\hD\ket{\rm vac}$, with
\begin{gather*}
\hS = \prod_{\vk} \hS_{\vk} = \prod_{\vk} \exp\lbc \frac{1}{2}
\lba \g_\vk\hcd_{\vk}\hcd_{-\vk} -
\g_\vk^*\hc_{\vk}\hc_{-\vk} \rba \rbc \, ,
\\
 \g_{\vk}+ \g_{-\vk} = 2 \ss_{\vk} e^{i\phi_{\vk}}  \, . 
\end{gather*}
Note that this automatically includes single-mode squeezing for the
condensate ($\vk=0$) mode, with squeeze parameter $\g_{0}$.  Our variational wavefunction for the
uniform interacting condensate is thus a squeezed coherent state for
the $\vk=0$ mode and a mixed-mode squeezed vacuum for each $\vk{\ne}0$
mode pair.

Minimization of the wavefunction locks the squeeze-parameter phases of
\emph{each} momentum-pair mode to twice the phase of the $\vk=0$
coherence parameter, i.e., $\phi_\vk = 2\th$ for \emph{all} $\vk$.  In
the following, we simply start with $\phi_\vk = 2\th$ to
avoid typing $(\phi_{\vk}-2\theta)$ arguments.

To determine the variational parameters, we need to minimize the
expectation value of the Hamiltonian \eqref{eq_WIBG-Hamilt}.  
Expectation values in the variational ground state $\ket{\rm sq1.gr}$ are 
calculated using the relations $\hDd\hc_\vk\hD = \hc_\vk +
\a\d_{\vk,0}$ and $\hSd\hc_\vk\hS =  \ch{\vk} \hc_{\vk} 
+ \sh{\vk} e^{i\phi_{\vk}} \hcd_{-\vk}$.  The required quantities are
$\langle\hat{N}_{\vk}\rangle = \langle\hcd_{\vk}\hc_{\vk}\rangle$ and
$\langle\hat{H}_{\rm int}\rangle$.
\begin{equation}   \label{eq_mode-occupancies}
\langle\hat{N}_{\vk}\rangle ~=~
  \sinh^2(\ss_\vk) + \Nc\d_{\vk,0}   \, ,
\end{equation}

and
\begin{multline*}
\langle\hat{H}_{\rm int}\rangle ~=~ 
\frac{U}{2V} \Nc^2    \\
+~ \frac{U}{2V} \Nc  
\lbb \sum_\vq\cosh(\ss_\vq)\sinh(\ss_\vq) 
+ 2\sum_\vq\sinh^2(\ss_\vq) \rbb          \\
~+~ \frac{U}{2V} \lbb \lbc\sum_\vq\cosh(\ss_\vq)\sinh(\ss_\vq)\rbc^2
+ 2 \lbc\sum_\vq\sinh^2(\ss_\vq)\rbc^2 \rbb   \, .
\end{multline*}
Here we have defined 
$\Nc ~=~ f^2 \lbc\cosh(2\ss_0) + \sinh(2\ss_0)\rbc$.
We can now minimize $\xpct{\hat{H}} = \sum_{\vk}
(\e_{\vk}-\mu)\xpct{\hat{N}_{\vk}} + \xpct{\hat{H}_{\rm int}}$ with
respect to $\Nc$ and $\ss_{\vk}$.  This yields
\begin{equation}   \label{mu_minimizn}
\mu ~=~  2Un + Um - 2U\nc     
\end{equation}
and
\begin{equation*}
\sinh(2\ss_\vk) = \frac{-Um}{\sqrt{(\e_\vk-\mu+2Un)^2-(Um)^2}}
\, .
\end{equation*}
Here we have defined $M = \sum_{\vq}\langle\hcd_\vq\hcd_{-\vq}\rangle
= \Nc+ \hf\sum_\vq \sinh(2\ss_\vq)\,$, and $m = M/V$.
We will show in Sec.~\ref{sect_spectrum} that the denominator appearing in $\sinh(2\ss_\vk)$
is the quasiparticle spectrum $E_{\vk}$, and that $\mu-2Un = -Um$.  Therefore
\begin{equation}  \label{eq_ss}
\sinh(2\ss_\vk) = \frac{-Um}{E_{\vk}} 
= \frac{-Um}{\sqrt{\e_{\vk}^2 + 2(Um)\e_{\vk}}}
\, .
\end{equation}

\subsection{Expectation values, variances and squeezing} 
\label{sect_xpctns-n-squeezings}

\emph{Condensate mode} ---
Expectation values of the bosonic operators are $\xpct{\hat{c}_0} =
\sqrt{\Nc}e^{i\th}$ and $\xpct{\hcd_0} =
\sqrt{\Nc}e^{-i\th}$. 
It is interesting to contrast this with Bogoliubov's mean-field
prescription, $\xpct{\hc_0} = \xpct{\hcd_0} = \sqrt{N_0}$.  Since $N_0
= \Nc + \sinh^2(\ss_0)$, the $\vk=0$ squeezing parameter measures the
deviation of our model from mean-field physics. This will be discussed
further in Sec.~\ref{sect_ss0-concl}.

Since we have used a squeezed coherent state for the condensate mode,
the expectation values and fluctuations of the quadrature operators
$\hat{X}_0 = \frac{1}{2}(\hc_0+\hcd_0)$ and $\hat{Y}_0 =
\frac{1}{2i}(\hc_0-\hcd_0)$ are identical to those for a squeezed
coherent state in quantum optics (Sec.~\ref{sect_quantopt},
Refs.~\cite{caves80, LoudonKnightReview87, loudon00_B}):
$\xpct{\hat{X}_0} = \Nc\cos\th$, $\xpct{\hat{Y}_0} = \Nc\sin\th$, and 
\begin{gather*}
\xpct{\d^2\hat{X}_0} ~=~
\frac{1}{4}\lbc e^{2\ss_0}\cos^2\th ~+~ e^{-2\ss_0}\sin^2\th \rbc 
\, ,  \\
\xpct{\d^2\hat{Y}_0}  ~=~
\frac{1}{4}\lbc e^{2\ss_0}\sin^2\th ~+~ e^{-2\ss_0}\cos^2\th \rbc
\, .      
\end{gather*}
The fluctuations along major (minor) axis directions are
$\hf{e}^{\pm\ss_0}$.

Eq.~\eqref{eq_ss} shows that the squeezing parameter $\ss_{\vk}$ is
negative, becomes large for small $\vk$, and diverges for $\vk = 0$.
This infrared divergence indicates that $\sinh(2\ss_0)$ scales as a
positive power of the system size, i.e.,
\[
\sinh(2\ss_0) = - \gamma_1 {N}^{\gamma_2} \quad \text{and} \quad  |\ss_0| \sim
\ord(\ln{N})
\, .  
\]
The $\gamma$'s can be extracted from finite-size considerations on
Eq.~\eqref{eq_ss}.  Noting that the lowest single-particle state in a
box of volume $V$ has energy $\e_0 = 3h^2/8\tilde{m}V^{2/3}$, we get
$\gamma_2 = 1/3$.  The same result can be obtained for a power-law
trap.  The exact number $\gamma_1$ is probably geometry-dependent.
However, ignoring factors like 2 and $\pi$ and the difference between
$n$ and $n_{\rm c}$, we find $\gamma_1 \sim \sqrt{n^{1/3}a}$; thus
\begin{equation}  \label{eq_sinh-2ss0}
\sinh(2\ss_0) \sim -\sqrt{n^{1/3}a}N^{1/3}\, .
\end{equation}
upto a factor of order 1.  In the true thermodynamic limit,
$\sqrt{n^{1/3}a}N^{1/3}\ra\infty$, so the variance profile is squeezed
infinitesimally thin in the radial direction.  The extension of the
variance profile in the phase direction,
$\sim\ord\lba[n^{1/3}a]^{1/4}N^{1/6}\rba$, although diverging, is
still infinitesimally small compared to the radial distance
($\sim\sqrt{N}$) of the state from the origin in the quadrature plane.
This is symptomatic of the fact that the squeezing has no
thermodynamic effects, which we show more explicitly in
Sec.~\ref{sect_ss0-concl}.

For finite-size systems, there is significant squeezing only for
$n^{1/3}aN^{2/3}\gg{1}$.  Note that this is the same condition that
decides whether the Thomas-Fermi approximation for a trapped
condensate is valid or not.

\begin{figure} 
\includegraphics[width=0.6\columnwidth]{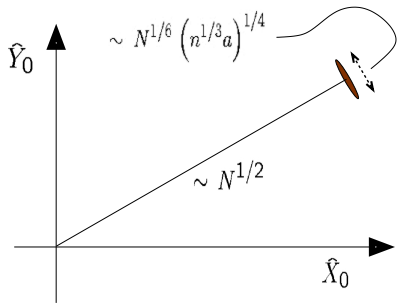}  
\caption{  \label{condensate-mode-squeezing} 
Squeezing of the condensate mode.  The variances are very much
exaggerated compared to the radial distance from the origin.  The
displayed variance to radial distance ratio, if drawn to scale, would
correspond to $N\sim\ord(10^3)$, in which case a weakly interacting
condensate ($an^{1/3}\lesssim{0.1}$) would be barely squeezed.
}
\end{figure}
%

%***  distribution of $N_0$.  Not a Poisson distribution.  ****

%**** fluctuation of $N_0$ ***  mr. navez's discussion needs to be
%     checked before referring to it.  My fluctuations don't seem to be
%     anomalous.  *****

\emph{Nonzero-momentum modes} ---  
The $\vk\ne0$ modes in the wavefunction $\ket{\rm sq1}$ have the
structure of two-mode squeezed vacua.  The operators have zero
expectation values, $\xpct{\hc_\vk} = \xpct{\hcd_\vk} = 0$.  There is
also no squeezing in quadrature operators defined within a single
mode: the usual $\hX_{\vk}$, $\hY_{\vk}$ have zero expectation values
and equal fluctuations.

Squeezing can be seen if one defines the mixed-mode operators 
\begin{gather*}
\hX_{\vk,-\vk} ~=~ \frac{1}{\sqrt{2}}(\hX_{\vk} +\hX_{-\vk})
~=~ \frac{1}{2\sqrt{2}} (\hc_\vk + \hc_{-\vk} + \hcd_\vk + \hcd_{-\vk})
\, ,   \\
\hY_{\vk,-\vk} ~=~ \frac{1}{\sqrt{2}}(\hY_{\vk} +\hY_{-\vk}) 
~=~ \frac{1}{2\sqrt{2}} (\hc_\vk + \hc_{-\vk} - \hcd_\vk -
\hcd_{-\vk})
\, .
\end{gather*}
These quadrature operators have zero expectation values and unequal
(squeezed) variances $\hf\ e^{\pm\ss_\vk}$.

The formalism of this section also allows us to calculate the
occupancies $N_{\vk} = \xpct{\hcd_{\vk}\hc_{\vk}}$ of the
non-condensate modes: $N_{\vk} = \sinh^2\ss_{\vk} = \hf\lbc
(\e_{\vk}+Um)/E_{\vk}-1\rbc$.  This is consistent upto mean field
order with standard treatments, e.g.,
Refs.~\cite{GlassgoldSauermann69a, ShiGriffin98}.

\section{Symmetry Breaking and Goldstone's Theorem}   \label{sect_U1}

We now investigate the symmetry-broken nature of the ground state of
the Bose gas.  The ground state has a particular phase, thus
spontaneously breaking a continuous $U(1)$ symmetry present in the
Hamiltonian.  Symmetry-broken ground states satisfy a Ward-Takahashi
identity reflecting the invariance of the ground-state energy under
shifts of the ground state by the symmetry operation in question.
According to Goldstone's theorem, a phase with broken continuous
symmetry should have a gapless mode.  The Ward identity and
gaplessness, both being consequences of the same phenomenon of
spontaneous symmetry breaking, are equivalent conditions and can
generally be derived from each other.
In the case of the Bose gas, the corresponding Ward identity is known
as the Hugenholtz-Pines (H-P) theorem \cite{HP59, HohMart64}.  It is
the condition for gaplessness as well as a consequence of the
invariance of the ground-state energy under shifts of the $U(1)$
phase.  The H-P theorem reads $\mu = \S_{11}(0,0) - \S_{12}(0,0)$,
where $\S_{11}(\vk,\o)$ and $\S_{12}(\vk,\o)$ are the normal and
anomalous self-energies.

In Sec.~\ref{sect_varyphase} we use the fact that the Hamiltonian has
a $U(1)$ symmetry while the ground state (and hence our variational
wavefunction) does not.  In Sec.~\ref{sect_spectrum} we calculate the
excitation spectrum by constructing a single-quasiparticle
wavefunction, and impose the requirement of gaplessness.  The two
considerations lead to the same condition for the variational
parameters, which is comforting in light of Goldstone's theorem.  The
condition should be equivalent to the H-P theorem.  In
Sec.~\ref{sect_ss0-concl} we compare our Ward identity with the
Hartree-Fock-Bogoliubov (mean-field) form of the H-P theorem, and
hence evaluate the importance and effects of condensate-mode
squeezing, $\ss_0$.

\subsection{$U(1)$ symmetry breaking}   \label{sect_varyphase}

In our variational wavefunction, the symmetry-broken nature of the
ground state appears as the definite phase of the coherence parameter
$\a = |\a|{\ }e^{i\th}$.  A shift of this phase would obviously change
the wavefunction, but should not affect the ground-state energy, since
the Hamiltonian is $U(1)$-invariant.  This requirement will give us
the Ward identity for our formalism corresponding to the H-P theorem.

We examine the transformation
\begin{gather*}
\a \ra \wt{\a} = \a e^{i\l} = f e^{i(\th+\l)}
\, , 
\quad \hat{D} \ra \wt{\hat{D}}  \, ,
\\
\g_{\vk} \ra \wt{\g_{\vk}} = \ss_{\vk} e^{i(2\th+2\l)} 
\, , 
\quad \hat{S} \ra \wt{\hat{S}}  
\, ,
\end{gather*}
so that the ground state is shifted, $\ket{\rm sq1.gr} \ra
\ket{\wt{\rm sq1.gr}} ~=~ (\wt{\hat{S}})^\dag\ (\wt{\hat{D}})^\dag \
\ket{\rm vac}$.  We consider infinitesimal $\l$, so that $\wt{\a}
\approx \a(1+i\l)$, and $\wt{\Nc} = \Nc(1+\l^2)$.

The shift in the thermodynamic potential is
\begin{multline*}
\d\langle\hat{H}\rangle 
= \bra{\wt{\rm sq1.gr}}\hat{H}\ket{\wt{\rm sq1.gr}}  
- \left<{\rm sq1.gr}|\hat{H}|{\rm sq1.gr}\right>  
\\
~=~ \l^2 \Nc \lbb -\mu + 2Un - U\nc \rbb   \,  .
\end{multline*}
We now use the requirement that the grand-canonical energy should not
be changed by a shift of the ground state phase, i.e.,
$\d\xpct{\hat{H}} = 0$.  Thus we get
\begin{equation}  \label{eq_mu_U1}
\mu = 2Un - U\nc 
~=~ 2Un - Un_0 + U \lbc \frac{\sinh^2(\ss_0)}{V} \rbc  \, \, .
\end{equation}
Within the variational formalism, this is our equivalent of the H-P
relation.

\subsection{Excitation spectrum}   \label{sect_spectrum}

We first construct a wavefunction for a Bose-gas ground state with a
single-quasiparticle excitation added on top of it:
\[
\ket{\rm sq1.ex(\vk_1)} = \hS\hD\, \hcd_{\vk_1} \ket{\rm vac}
\; .
\]
The idea is that, since the squeeze operator $\hS$ represents
interaction effects in the present formalism, the particle creation
operator $\hat{c}_{\vk_1}^\dag$ should produce a Bogoliubov
quasiparticle when used in conjunction with $\hS$ .

One can evaluate matrix elements of $\hat{N}_{\vk}$ and $\hat{H}_{\rm
int}$ in the state $\ket{\rm sq1.ex(\vk_1)}$ just as was done in the
ground state $\ket{\rm sq1.gr}$.  The calculation is lengthier but
straightforward.  Calculating $\xpct{\hat{H}}$, one now obtains the
excitation spectrum as the (grand-canonical) energy of the new state
with respect to the ground state.  
\begin{multline*}
E_{\vk} ~=~ 
\\
\bra{\rm sq1.ex(\vk_1)}\hat{H}\ket{\rm sq1.ex(\vk_1)} 
~-~ \bra{\rm sq1.gr}\hat{H}\ket{\rm sq1.gr}   \\
 =~ \cosh(2\ss_{\vk_1}) \lbc \e_{\vk_1}-\mu + 2Un \rbc +~
\sinh(2\ss_{\vk_1})\cdot Um   \\
~=~ \lbc(\e_{\vk_1}-\mu + 2Un)^2 - (Um)^2 \rbc ^{1/2} 
\end{multline*}
For the spectrum to be gapless, one requires $\mu-2Un = \pm Um$.  The
positive sign is inconsistent (resulting in $\nc =0$).  Therefore, we
have
\begin{equation}  \label{mu_gapless}
\mu = 2Un  - Um
\end{equation}
Taken together with Eq.~\eqref{mu_minimizn}, this is indeed
identical to the condition \eqref{eq_mu_U1} obtained from consideration
of $U(1)$ symmetry breaking, as expected. 

The spectrum we have is thus $E_{\vk} = \sqrt{\e_{\vk}^2
+2(U\nc)\e_{\vk}} = \sqrt{\e_{\vk}^2+2(Um)\e_{\vk}}$\,, which may
be contrasted with the Bogoliubov spectrum $E_{\vk} =
\sqrt{\e_{\vk}^2 +2(Un_0)\e_{\vk}}$\,.

\subsection{Inferences on condensate mode squeezing}   \label{sect_ss0-concl}

In the thermodynamic limit, $N\ra\infty$, Eq.~\eqref{eq_sinh-2ss0}
implies $\sinh^2(\ss_0) \approx -\hf\sinh(2\ss_0) \sim
\sqrt{n^{1/3}a}N^{1/3}$.  Thus, in our Ward identity
Eq.~\eqref{eq_mu_U1}, the contribution from $\ss_o$ vanishes as
$N^{-2/3}$ for macroscopic systems.  The effect of condensate-mode
squeezing on other thermodynamic quantities and equations similarly
vanishes in the $N\ra\infty$, $V\ra\infty$ limit, since $\ss_0$
generally appears as $\sinh^2(\ss_0)$ or $\sinh(2\ss_0)$ in equations
involving extensive quantities.

At the mean field Hartree-Fock-Bogoliubov (HFB) level, $\S_{11} = 2Un$
and $\S_{12} = Un_0$, so that the H-P theorem is $\mu = 2Un - Un_0$.
Comparing with our form $\mu = 2Un-\nc = 2Un-m$, we conclude that the
$\hS\hD\ket{\rm vac}$ formalism reduces to the mean-field HFB results
if $N_0 = \Nc = M$.  Noting from Eq.~\eqref{eq_mode-occupancies} that
\begin{equation}  \label{N0Nc}
N_0 = \xpct{\hcd_0\hc_0} = \Nc + \sinh^2(\ss_0)  \, ,
\end{equation}
the condition for our formalism to be restricted to mean-field physics
is $\ss_0 = 0$.  The $\vk=0$ squeezing parameter is thus a measure of
the deviation of the formalism from HFB physics.
The point is further emphasized by rewriting Eq.~\eqref{N0Nc} as
$\xpct{\hcd_0\hc_0} = \xpct{\hcd_0} \xpct{\hc_0} + \sinh^2(\ss_0)$,
which shows that $\sinh^2(\ss_0)$ acts as a correction to mean-field
type decomposition.
The argument can be inverted to state that, at mean field level, the
weakly interacting $T=0$ Bose gas has no squeezing in the
zero-momentum mode.  Since $\ss_0 = 0$ at mean field level, the $\vk=0$
squeezing must come from beyond mean field.  

It may seem tempting to try to identify which diagrams contribute to
$\ss_0 = 0$, i.e., to identify contributions to $\S_{12}$ or $\S_{11}$
of the form $U\sqrt{n^{1/3}a}N^{-2/3}$.  Note however that these would
be non-extensive contributions, which are (not surprisingly) not
readily found in the literature.

Note that, in contrast to the $\vk=0$ mode squeezing, the nonzero
$\pm\vk$ mixed-mode squeezing in the $\vk{\ne}0$ modes is present at
mean field level already.

\section{Squeezing in ``Fixed-$N$'' Excitations}   \label{sect_AER}

In this Section, we will introduce and study a second variational
formulation of the interacting Bose condensate in order to give a more
physical interpretation of the squeezing of the nonzero-momentum modes.
The formalism will be based on the the bosonic operators introduced by
A.~E.~Ruckenstein in Ref.~\cite{fixed-N:AER00}.

\subsection{Bosonic fields and excitation Hamiltonian}   
\label{sect_AER:b+Hx}

Ref.~\cite{fixed-N:AER00} presents a current algebra approach to
formulating a number-conserving description of the Bose condensate.  The
Hamiltonian is written in terms of density and current operators
$\hat{n}(\vr) = \hcd(\vr)\hc(\vr)$ and $\hat{\bf j} =
-\tfrac{i}{2m}[\hcd(\vr){\bf\nabla}\hc(\vr) -
{\bf\nabla}\hcd(\vr)\hc(\vr)]$.

The density fluctuation operator $\heta$, defined as $\heta(\vr) =
\hat{n}(\vr)-\xpct{\hat{n}(\vr)} = \hat{n}(\vr)- \nG(\vr)$, and the
phase operator $\hat{\phi}$, defined by $\hat{\bf j} =
\frac{\nG(\vr)}{m} {\bf\nabla}\hat{\phi}(\vr)$, are canonically
conjugate.  Defining the linear combinations $\hb(\vr)$,
$\hbd(\vr)$ with 
\[
\hb(\vr)  = \frac{1}{2\sqrt{\nG(\vr)}} \lbc \heta(\vr) +
2i\nG(\vr)\hat{\phi}(\vr)    \rbc  \, ,
\]
the Hamiltonian in \cite{fixed-N:AER00} takes the form 
\[
\hat{H} ~=~ E_{\rm GS} \lbc\nG(\vr)\rbc  ~+~ 
\hat{H}_X \lbc \nG(\vr),\hb(\vr),\hbd(\vr) \rbc \, .
\]
$E_{\rm GS}$ describes the mean-field ground state, and minimizing
this functional gives an equivalent of the Gross-Pitaevskii equation
which determines $\nG(\vr)$.  In this Article we concentrate on the
uniform case, $\nG(\vr)=\nG$.  We are more interested in the
excitation Hamiltonian $\hat{H}_X$ which describes the low-lying,
large-lengthscale excitations. In momentum space, $\hat{H}_X$ reads
\begin{multline}   \label{eq_AER:Hx}
\hat{H}_X ~=~ \sum_{\vk{\ne}0} \e_\vk\ \hbd_\vk\hb_\vk 
\\
~+~ \hf{U}\nG
\sum_{\vk{\ne}0} \lba \hbd_\vk\hbd_{-\vk} + \hb_\vk\hb_{-\vk} + 2\
\hbd_\vk\hb_\vk \rba  \, ,
\end{multline}
modulo an additive constant.  Here $\hb_\vk = \int_\vr \hb(\vr)
e^{-i\vk\cdot\vr}$.  

The Hamiltonian \eqref{eq_AER:Hx} looks identical to that derived by
Bogoliubov.  However, the operators in the Bogoliubov Hamiltonian are
the original bosonic operators $\hc$, $\hcd$, rather than the peculiar
bosons $\hb$, $\hbd$, that we have here.  The interpretation is very
different; the Bogoliubov picture involves an order parameter and
$\pm\vk$ pairs can appear from or disappear into the condensate, while
in the fixed-$N$ picture, there is no order parameter.  $\hat{H}_X$
should not be regarded as a quasiparticle Hamiltonian, but rather as
the Hamiltonian describing low-lying \emph{density and current
oscillations} of the system at a fixed total particle number.  It is
then no surprise that Eq.~\eqref{eq_AER:Hx} does not conserve the
number of bosons, $\int_\vr\hbd(\vr)\hb(\vr)$.  Our reason for using
this formalism is that the $\hb$ bosons can be interpreted in terms of
density fluctuation and phase operators.

Introducing Fourier transforms of the density fluctuation and phase
operators, $\heta_\vk = \int_\vr \heta(\vr) e^{-i\vk\cdot\vr}$,
and $\hph_\vk = \int_\vr \hph(\vr) e^{-i\vk\cdot\vr}$.  
we can express mixed-mode hermitian quadrature operators as 
\begin{equation} \label{eq_AER:XY}
\begin{split}
\hX_{\vk,-\vk} = \frac{1}{4}\lba \hb_{\vk} + \hb_{-\vk} + \hbd_{\vk} +
\hbd_{-\vk} \rba  
= \frac{1}{4\nG} \lba \heta_{\vk} + \hetad_{\vk} \rba  \, ,
\\
\hY_{\vk,-\vk} = \frac{1}{4i}\lba \hb_{\vk} + \hb_{-\vk} - \hbd_{\vk} -
\hbd_{-\vk} \rba 
= \frac{\nG}{2} \lba \hph_{\vk} + \hphd_{\vk} \rba  \, .
\end{split}
\end{equation}
Note that $\hX_{\vk,-\vk}$ and $\hY_{\vk,-\vk}$ here are different
from the quadrature operators defined in
Sec.~\ref{sect_xpctns-n-squeezings} because the bosons $\hb$, $\hbd$
have different meanings from $\hc$, $\hcd$.

\subsection{Variational treatment, squeezing}   
\label{sect_AER:variation}

Let us define the reference state $\ket{\rm ref}$ as the vacuum for
the $\hb_\vk$ bosons.  We now introduce the following wavefunction as
a variational state for the system:
\[
\ket{\rm sq2} = \prod_{\vk{\ne}0}  \exp\lbc 
\g_\vk\hbd_{\vk}\hbd_{-\vk} - \g_{\vk}^*\hb_{\vk}\hb_{-\vk}
\rbc    \ket{\rm ref}
= \hS \ket{\rm ref}   \, ,
\]
with the usual $\g_{\vk} + \g_{-\vk} = 2\ss_{\vk}\,e^{i\phi_\vk}$.

The state $\ket{\rm ref}$ itself is determined from the
$E_{\rm GS}$ part of the theory.
The expectation values in our variational state $\ket{\rm sq2}$ are
$\xpct{\hbd_{\vk}\hb_{\vk}} = \sinh^2(\ss_\vk)$ and
$\xpct{\hbd_{\vk}\hbd_{-\vk}} = \hf\sinh(2\ss_\vk) e^{-i\phi_\vk}$.
We will minimize $\xpct{H_X}$, not $\xpct{H_X - \mu\hat{N}}$.  This is
because the number of excitations are not conserved, and $\mu$ is
determined in the $E_{\rm GS}$ part of the theory.
Minimization leads to  $\phi_{\vk} = 0$, and
\[
\sinh(2\ss_\vk) = \frac{-U\nG}{\sqrt{\e_{\vk}^2 + 2U\nG\e_{\vk}}} 
\, .
\]

One can also calculate the excitation spectrum from this alternate
variational procedure.  As in Sec.~\ref{sect_spectrum}, we can
construct the excited state $\ket{\rm sq2.ex(\vp)} = \hS{\
}\hbd_\vp\ket{\rm ref}$.  Using expectation values in states $\ket{\rm
sq2.gr}$ and $\ket{\rm sq2.ex(\vp)}$, the dispersion relation is found
to be $E(\vp) = \sqrt{ \e_\vp^2 + 2\ (U\nG)\ \e_\vp}$.  This is the
Bogoliubov spectrum, assuming $\nG = n_0$.  This demonstrates that our
new variational formulation captures the physics of the weakly
interacting Bose gas at least up to mean field level.  We are
therefore justified in using the formalism based on the state
$\ket{\rm sq2}$ to draw conclusions about the $\vk{\ne}0$ mixed-mode
squeezing.

\begin{figure} 
\begin{center} 
  \includegraphics[width=0.25\columnwidth]{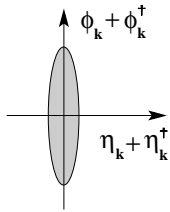}  
\caption[Squeezing of nonzero-momentum modes, in terms of physical
variables.]{
\label{fig_nonzero-mode-squeezing} 
Squeezing of nonzero-momentum modes, in terms of ``physical''
variables.  The variables are Fourier transforms of density
fluctuation and phase, in units of $\nG$ and $\nG^{-1}$ respectively.
The density fluctuation variables are squeezed.  The squeezing shown
here is very moderate, i.e., a large-momentum mode or a condensate
with small number of particles.}
\end{center}
\end{figure}

Just as in Sec.~\ref{sect_xpctns-n-squeezings} for the state $\ket{\rm
sq1}$, and in Sec.~\ref{sect_quantopt} for a general two-mode squeezed
vacuum, the state $\ket{\rm sq2}$ displays squeezing in the plane of
mixed-mode quadrature operators $\hX_{\vk,-\vk}$ and $\hY_{\vk,-\vk}$.
However, at this stage the relevant quadrature operators are
physically meaningful:
$\hX_{\vk,-\vk} = \frac{1}{4\nG} \lba \heta_{\vk} + \hetad_{\vk} \rba$
and
$\hY_{\vk,-\vk} = \frac{\nG}{2} \lba \hph_{\vk} + \hphd_{\vk} \rba$,
as defined in Eq.~\eqref{eq_AER:XY}.
Since $\ss_\vk$ is negative, squeezing is along the $\hX_{\vk,-\vk}$
direction (Fig.~\ref{fig_nonzero-mode-squeezing}).  The squeezing is
larger for lower momentum.

Thus our study of the alternate variational formulation, in terms of
the bosons of the ``fixed-$N$'' theory \cite{fixed-N:AER00}, has
allowed us to express a well-known squeezing phenomenon in terms of
variables that have the very physical meaning of density fluctuations
and phases, albeit in momentum space.  This may be regarded as a new
formulation of the old idea, attributed to Feynman
\cite{feynman:statmech_B, tc-feynman53}, that in a repulsive Bose
condensate, density fluctuations should be suppressed (or in modern
language, squeezed).

\section{Discussion} \label{sect_concl}

In summary, we have addressed in detail the issue of squeezing in
various modes of the ground state of a uniform condensate, using two
different variational wavefunctions.

For squeezing in the condensate mode, we have presented a clear
analysis of the scaling of squeeze parameter $\ss_0$ with system size,
using our first wavefunction $\ket{\rm sq1}$, resulting in the scaling
relation $e^{2\ss_0} \sim -\sqrt{n^{1/3}a}N^{1/3}$.  This leads to the
conclusion that while the ground state is indeed squeezed (with
the uncertainty profile distortion even diverging for $N\ra\infty$)
the squeeze parameter nevertheless has no thermodynamic effects.  For
finite-size systems, such as condensates in traps, we have identified
that the Thomas-Fermi regime ($an^{1/3} \gg N^{-2/3}$) is the
interaction regime where one expects to see appreciable squeezing of
the ground state.

Our second wavefunction $\ket{\rm sq2}$ is devised specifically to
address the issue of pair squeezing in the non-condensate
opposite-momenta mode pairs.  Using results from one of the
$U(1)$-invariant formulations of the Bose condensate
\cite{fixed-N:AER00}, we have provided an interpretation of this pair
squeezing in terms of variables representing density and phase
excitations.

We now make contact with relevant results in the
literature.  It is worth pointing out that our treatment of
gaplessness, where imposing the Hugenholtz-Pines theorem leads to the
condition $m=\nc$, is actually equivalent to the Popov approximation
\cite{popov:FICE_B, griffin_ConservingGapless96, ShiGriffin98} where
anomalous pair correlation functions ($\tilde{m}$ in
Ref.~\cite{griffin_ConservingGapless96}) are neglected.  This is a
simple and direct way to implement gaplessness.  In
Ref.~\cite{navez98}, a more involved procedure for satisfying the
Hugenholtz-Pines theorem leads to a \emph{macroscopic} condensate-mode
squeezing.  However this contradicts the scaling relationship
$e^{2\ss_0} \sim -N^{1/3}$ that we have derived here directly from the
minimization of variational parameters.  Other scattered previous
discussions of squeezing in the condensate mode
\cite{SolomonFengPenna99, DunninghamCollettWalls98,
RS-Choi-New-Burnet02:quantumstate, GlassgoldSauermann69a,
GlassgoldSauermann69b} have not addressed clearly the role of this
squeezing in the thermodynamic limit.  Finally, concerning the density
fluctuation operators borrowed from Ref.~\cite{fixed-N:AER00}, we note
that similar operators have appeared in other fixed-$N$ formulations
of the condensate ground state, e.g., in
Ref.~\cite{fixed-N:gardiner97}.

We end by pointing out some open problems.  

Our results on the condensate-mode squeezing prompts questions about
the presence of squeezing in \emph{trapped} condensates.  Study of the
quantum state of trapped condensates, either experimentally through
quantum state tomography methods or theoretically, is essential for
verifying our finite-size scaling relation $e^{2\ss_0} \sim
-\sqrt{n^{1/3}a}N^{1/3}$.  Refs.~\cite{DunninghamCollettWalls98,
RS-Choi-New-Burnet02:quantumstate} have reported $Q$-function and
Wigner function calculations of the condensate quantum state, showing
squeezing in a number of cases.  However, no systematic analysis of
the $N$-dependence or interaction-dependence of the squeezing
parameter is available.  

Another question related to the quantum state of the condensate mode
is the possibility of non-classical features other than squeezing.  A
whole number of quantum states are studied in quantum optics (Fock,
thermal, squeezed Fock, etc.) and it is intriguing to ask if, for
example, using a squeezed Fock state instead of a squeezed coherent
state for the $\vk=0$ mode would gain us complementary insight.
%
%% (The 1959 work of Ref.~\cite{GirardeauArnow59} uses a state similar to
%% a squeezed Fock state.)
%
Also, other quantum states might be helpful in exploring $\vk\ne{0}$
physics beyond the mean-field level physics we have extracted here for
the non-condensate modes.  Inclusion of other quantum-state features
might also be fruitful for a variational description of
finite-temperature, two-dimensional, or trapped Bose gases.

%% A known difficulty with the formalism of Secs.~\ref{sect_formalism}
%% and \ref{sect_U1} concerns the Gavoret-Nozierres theorem
%% \cite{GavoretNozieres64, HohMart64, ShiGriffin98}.  It became clear
%% during the work of Ref.~\cite{navez98} that this theorem is not
%% straightforwardly satisfied beyond leading order
%% \cite{navez_personal-comm}.  This issue remains unresolved to the best
%% of the present authors' knowledge.

A \emph{real-space} variational procedure using wavefunctions of
Jastrow form has often been employed to describe interacting Bose
condensates \cite{BDJ:SimWooBuchler70_PRL, BDJ:LeeWong75,
feenberg69_B}.  A natural question is the relation to our variational
description.  Presumably, the success of the so-called
Bijl-Dingle-Jastrow wavefunction is due to its correctly capturing
correlations such as those we have discussed in terms of squeezing.
However, it remains unclear how to extract from real-space Jastrow
wavefunctions the momentum-space squeezing parameters of the type
included in our $\hS\hD$ state $\ket{\rm sq1}$.

%%%%%%%%%%%%%%%%%%%%%%%%%%%%%%%%%%%%%%%%%%%%%%%%%%%%%%%%%%%%%%%%%%

\acknowledgments

Helpful conversations with Morrel H.~Cohen, Alan Griffin, Patrick
Navez, and Henk Stoof are gratefully acknowledged.  MH was funded by
the Nederlandse Organisatie voor Wetenschaplijk Onderzoek (NWO).

%%%%%%%%%%%%%%%%%%%%%%%%%%%%%%%%%%%%%%%%%%%%%%%%%%%%%%%%%%%%%%%%%%

%%%%%%%%%%%%%%%%%%%%%%%%%%%%%%%%%%%%%%%%%%%%%%%%%%%%%%%%%%%%%%%%%%

%\bibliography{squBEC_paper_f05,NewEraBECthy,Tc,quantoptics,books,BECbasics,thesis-chap1,2Dbose,AtomicGasExpts}
    
%\bibliographystyle{prsty}

%%%%%%%%%%%%%%%%%%%%%%%%%%%%%%%%%%%%%%%%%%%%%%%%%%%%%%%%%%%%%%%%%%

\end{document}